\documentclass[10pt,journal]{IEEEtran} 
\usepackage[utf8]{inputenc}
\usepackage[T1]{fontenc}
\usepackage{amssymb,amsmath}
\usepackage{cite}
\usepackage{psfrag}
\usepackage{url}
\usepackage[absolute,overlay]{textpos}

\usepackage{pgf}
\usepackage[export]{adjustbox}
\usepackage{bm}
\usepackage{bbm}
\usepackage{booktabs}
\usepackage{url}
\usepackage{upgreek}
\usepackage{verbatimbox}
\usepackage{algorithm}
\usepackage{algorithmic}
\usepackage{enumitem}

\newcommand{\appropto}{\mathrel{\vcenter{
			\offinterlineskip\halign{\hfil$##$\cr
				\propto\cr\noalign{\kern2pt}\sim\cr\noalign{\kern-2pt}}}}}
\usepackage{amsthm}

\usepackage{colortbl}

\begin{document}
	\title{Dynamic RF Combining for  Multi-Antenna Ambient Energy Harvesting}		
	\author{ Onel L. A. López,~\IEEEmembership{Member,~IEEE,}
	Bruno Clerckx,~\IEEEmembership{Senior Member,~IEEE},\\
	and Matti Latva-aho,~\IEEEmembership{Senior Member,~IEEE}
	\thanks{Onel López and Matti Latva-aho are  with  the Centre for Wireless Communications University of Oulu, Finland, e-mails: \{Onel.AlcarazLopez, Matti.Latva-aho\}@oulu.fi.}
	\thanks{Bruno Clerckx is with Department of Electrical and Electronic Engineering, Imperial College London, U.K, 	e-mail: b.clerckx@imperial.ac.uk.}
	\thanks{This research has been financially supported by Academy of Finland through 6Genesis Flagship (Grant no. 318927).} 
	\thanks{This work has been accepted for publication in IEEE Wireless Communications Letters. © 2021 IEEE. Personal use of this material is permitted. Permission from IEEE must be obtained for all other uses, in any current or future media, including reprinting/republishing this material for advertising or promotional purposes, creating new collective works, for resale or redistribution to servers or lists, or reuse of any copyrighted component of this work in other works.}
	} 	

	\maketitle
\begin{abstract}
    Ambient radio frequency (RF) energy harvesting (EH) technology is  key to realize self-sustainable, always-on, low-power, massive Internet of Things  networks. Typically, rigid (non-adaptable to channel fluctuations) multi-antenna receive architectures are proposed to support reliable EH operation. Herein, we introduce a dynamic RF combining architecture for ambient RF EH use cases, and exemplify the attainable performance gains via three simple phase shifts' exploration mechanisms, namely, brute force (BF), sequential testing (ST) and codebook based (CB). Among the proposed mechanisms, BF demands the highest power consumption, while CB requires the highest-resolution phase shifters, thus tipping the scales in favor of ST. Finally, we show that the performance gains of ST over a rigid RF combining scheme increase with the number of receive antennas and energy transmitters' deployment density.
\end{abstract}
	\begin{IEEEkeywords}
		ambient RF energy harvesting, multiple antennas, phase shifts,  power consumption, dynamic RF combining.
	\end{IEEEkeywords}
	\IEEEpeerreviewmaketitle
\section{Introduction}\label{intro}
Ambient energy harvesting (EH) technologies are  key enablers of self-sustainable always-on  networks \cite{Lopez.2021,ZhangGrajal.2020,Liang.2020}. Among the variety of ambient EH sources, e.g., vibration, heat and light, ambient radio frequency (RF) EH is particularly attractive for powering massive Internet of Things (IoT) deployments. The reasons are: i) RF EH modules can be easily incorporated into small-form factor IoT nodes \cite{Lopez.2021}, ii) although in general the energy harvestable from ambient RF signals is low, it may suffice to support the operation of the new-generation of  ultra-low power sensors and passive IoT actuators (the so-called `Internet of Tiny Things' \cite{ZhangGrajal.2020}), and iii) due to a denser coexistence of multiple wireless services \cite{Liang.2020},  RF energy availability is superior in urban/suburban environments, where massive IoT networks are mostly deployed.

\begin{figure}[t!]
\centering
\includegraphics[width=0.49\textwidth]{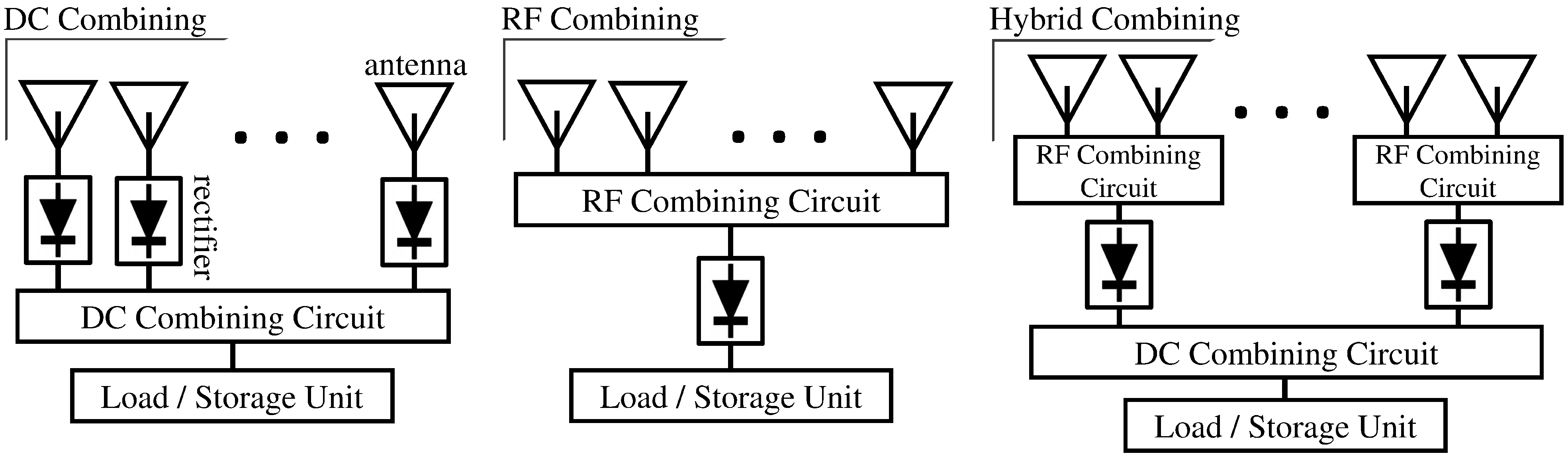}
\vspace{-6mm}
\caption{EH architecture when using a) DC combining (left), b) RF combining (middle), and c) hybrid combining (right).}
\label{Fig1}
\vspace{-4mm}
\end{figure}

Typically, multiple antennas are needed to provide enough energy for a reliable EH  operation \cite{Lee.2017,Wu.2017,Olgun.2011,Shen.2020}. The basic three EH multi-antenna architectures are: i) DC combining (Fig.~\ref{Fig1}a), where each antenna branch incorporates its own rectifier to separately harvest  power, ii) RF combining (Fig.~\ref{Fig1}b), where all the antennas are arranged to channel the RF power to a single rectifier, and ii) hybrid combining  (Fig.~\ref{Fig1}c). DC combining provides broader beamwidth than RF combining, thus, allowing EH from a broader range of incident directions. However, the rectifiers must individually rectify low RF powers,  resulting in low conversion efficiency. RF combining, on the other hand, feeds the combined signal to a rectifier, which therefore operates on higher input power levels and achieves higher conversion efficiencies \cite{Olgun.2011,Shen.2020}. Hybrid architectures aim at bringing together the large beamwidth and higher conversion efficiencies of previous standalone approaches.

The state-of-the-art research and prototyping considers DC combining or rigid (non-tunable) RF/hybrid architectures for ambient RF EH mainly because of the terminals' ultra-low-power consumption requirements. However, with the development of ultra-low-power integrated circuit designs and ultrathin, flexible energy harvesters with improved power efficiency \cite{ZhangGrajal.2020}, dynamic  implementations may be at hand. Specifically, a dynamic RF combining circuit may conveniently tune the phase shifts at each antenna branch as to maximize the harvested energy. For dedicated closed-loop RF wireless energy transfer (WET), it has been shown in \cite{Shen.2021,ShenClerckx.2021} that dynamic RF combining can provide substantial performance gains (more than double the performance) over DC combining when exploiting full channel state information (CSI) by leveraging the rectenna nonlinearity more efficiently.

However, ambient RF EH setups impose critical challenges for the design/adoption of dynamic RF combining, specifically: i) EH frequency bandwidths are usually broad, thus, channels are frequency-selective, ii) multiple RF signals may potentially coexist, and iii) phase/magnitude CSI cannot be acquired (at least in a standard manner)\footnote{This is also a characteristic of open-loop WET systems, for which our proposed architecture and mechanisms would be beneficial as well. However, since WET systems operate usually in closed-loop to boost the EH efficiency, our motivation and discussions are focused on ambient RF EH.}. In this letter, we propose a dynamic multi-antenna RF combining architecture and accompanying phase shifts' exploration mechanisms to efficiently support ambient low-power RF EH. Specifically, our contributions are three-fold: i) we introduce a dynamic RF combining architecture for ambient RF EH use cases addressing aforementioned challenges, ii) we identify three simple phase shifts' exploration mechanisms, namely, brute force (BF), sequential testing (ST) and codebook based (CB) and investigate their performance, and iii) we consider the extra energy consumption required by the proposed mechanisms and performance degradation due to the phase shifters' finite resolution and  insertion losses. We show that the performance of the proposed mechanisms approaches asymptotically the optimum as the corresponding energy consumption becomes small with respect to the energy to be harvested. Among the proposed mechanisms, BF demands the highest power consumption, while CB requires the highest-resolution phase shifters, thus tipping the scales in favor of ST. Finally, we show that the performance gains of ST over a rigid combining (RC) scheme increase with the number of receive antennas and energy transmitters' deployment density.
\section{System Model}\label{system}
Consider an IoT device equipped with $M$ antennas and scavenging ambient RF energy from an arbitrary set $\mathcal{S}$ of nearby transmitters operating in the frequency band of interest (antenna bandwidth). The device's circuitry is configured to scavenge energy via an RF combining architecture as illustrated in Fig.~\ref{Fig1}b. Let's denote by $\mathbf{r}(t)\in\mathbb{C}^{M\times 1}$ (with entries $r_j(t),\ j=1,\cdots,M$) the RF signal captured by the EH antennas and impinging the RF power combining circuit
at time instant $t$. Observe that  each receive antenna is connected to a common RF combining circuit feeding a single rectifying circuit. The RF combining circuit is  composed of:
\begin{itemize}
	\item an array of phase shifters. Let us denote by $\delta_j\in(0,1)$ the insertion loss of the $j-$th phase shifter \cite{Ocera.2005}, and by $\theta_j\in\Theta\subseteq [0,2\pi]$ its configured phase shift. We assume that $\theta_j$ is uniform along the antenna bandwidth \cite{Ellinger.2008};
	\item a passive power combiner, thus, the output power is equal or less than the input power \cite{Shen.2021,ShenClerckx.2021}.
\end{itemize}
The harvested energy in a time interval $[t_1,t_2]$ is given by
\begin{align}
   E_{t_1}^{t_2} (\bm{\theta}) &= \int_{t_1}^{t_2}g\bigg(\frac{1}{ \sqrt{M}}\sum_{j=1}^{M}r_j(t)\sqrt{\delta_j}e^{\mathbbm{i}\theta_j}\bigg)\mathrm{d}t, \label{Et1t2}
\end{align} 
where $\bm{\theta}=[\theta_1,\theta_2,\cdots,\theta_M]^T$, and $g:\ \mathbb{C}\rightarrow\mathbb{R}^+$ models the relation between the RF signal (at the output of the power combiner) and harvested DC power. Here, the term $1/\sqrt{M}$ mimics the passive power combiner implementation.

Observe that $\mathbf{r}(t)$ is specified in the time domain, thus, it can have any kind of frequency response, which is determined by that of the receive antennas (including antenna bandwidth), transmit signals' waveform and diversity, and channel impairments. Moreover, the EH model used here is generic and can capture frequency-dependent fluctuations. Interestingly, the EH performance can be tuned by  $\bm{\theta}$. However, the state-of-the-art designs of RF combining circuits for ambient RF EH applications are rigid, e.g., \cite{Olgun.2011,Shen.2020,Lee.2017}, which means that $\bm{\theta}$ is pre-configured and remains unchangeable to favor low-complexity/power implementations. In this paper, we propose and discuss a low-complex dynamic RF combining architecture for ambient RF EH use cases. Our proposal can serve also to realize dynamic hybrid architectures by modifying the traditionally rigid RF combining circuit components. 
\section{Preliminaries on Dynamic RF Combining}\label{design}
\begin{figure}[t!]
	\centering
	\includegraphics[width=0.5\textwidth]{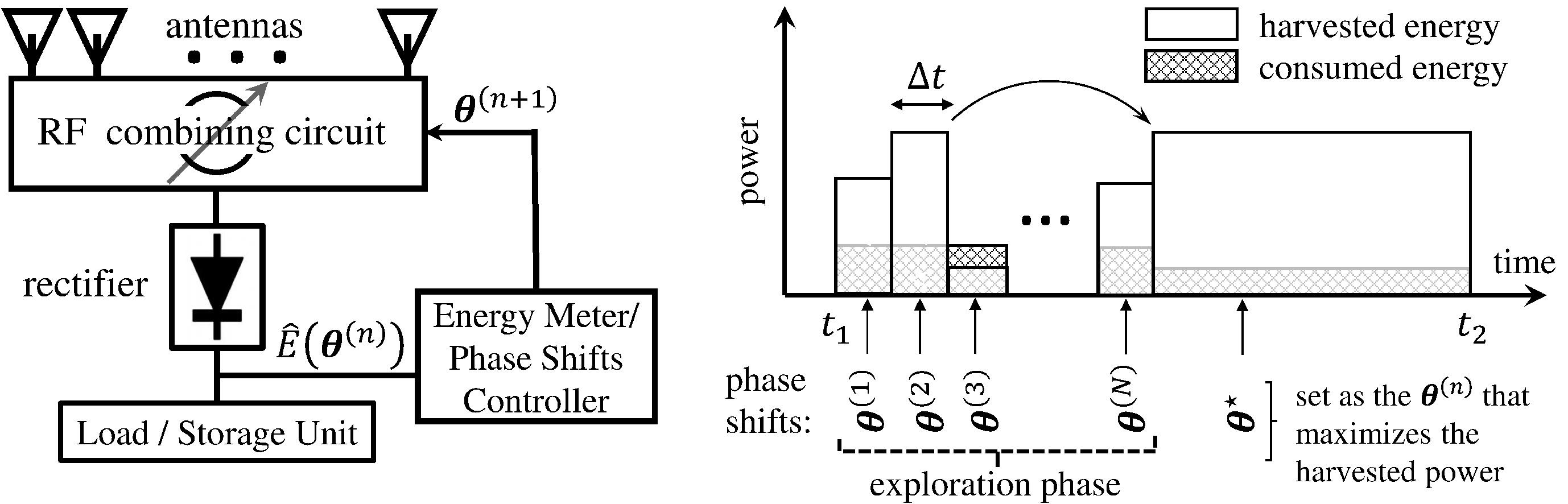}
	\vspace{-4mm}
	\caption{Dynamic RF combining: a) architecture (left), and b) operation basics (right). In b), and as an example, $\bm{\theta}^{(2)}$ is assumed to provide the greatest harvested power, thus,  $\bm{\theta}^{\star}\leftarrow \bm{\theta}^{(2)}$.}
	\label{Fig2}
	\vspace{-2mm}
\end{figure}

The optimum phase shift configuration is given by  $\bm{\theta}^\text{opt}=\arg\max_{\bm{\theta}\in\Theta}E_{t_1}^{t_2}(\bm{\theta})$.   
In general, the optimum design requires some specific knowledge of $\mathbf{r}(t)$, e.g., experienced channel realizations, which is not available in ambient RF EH scenarios. 
Still, phase shifts can be configured by exploiting energy measurement feedback from an energy meter at the output of the rectifier as shown in Fig.~\ref{Fig2}a. A somewhat similar approach is exploited \cite{Xu.2014,Xu.2016} but for the case of dedicated WET, where the energy measurements at a single-antenna or DC-combining multi-antenna EH receiver are feedbacked to control the transmit processing. For each measurement feedback, the technique of analytic center cutting plane method (ACCPM) is used to iteratively restrict the set on which the optimum beamformers lie until eventual convergence. This technique is undoubtedly appealing for a dedicated WET scenario, where the energy transmitter carries out the energy-consuming optimization operations, but much less for our considered setup where such operations need to be handled by very-low cost/power devices\footnote{The complexity for solving the ACCPM optimization problems increases exponentially with the number of feedback intervals since the set of added cutting planes becomes large. Authors in \cite{Xu.2016} rely on pruning irrelevant cutting planes to prevent a complexity increase with the number of feedback samples, but still this method remains unaffordable for ambient RF EH applications.}. 

Because of the ultra-low energy levels that are harvestable from ambient RF signals, any circuit design proposal must consider the performance degradation coming from the power consumption of the circuitry elements and insertion losses. We delve into these issues in the following.
\subsection{Net Harvested Energy}
The operation basics of the general dynamic RF combining architecture is illustrated in Fig.~\ref{Fig2}b.
Let us denote by $\Delta t\ll t_2-t_1$ the duration of a single measurement time interval out of the $N$ that are carried out. Then, the net harvested energy under any dynamic RF combining scheme can be written as $E_\text{net}=E_\text{abs}-E_c$ with
\begin{align}
E_\text{abs} &= \sum_{n=1}^{N}E_{t_1+(n-1)\Delta t}^{t_1+n\Delta t}(\bm{\theta}^{(n)})+E_{t_1+N\Delta t}^{t_2}(\bm{\theta}^{\star}),\label{Eabs}\\
E_c&=E_\text{phase}+N \Delta t P_\text{meter}, 
\end{align}
where $E_\text{abs}$ is the absolute harvested energy, and $E_c$ accounts the energy consumed by the dynamic phase shifting circuitry ($E_\text{phase}$), and the metering circuit. The latter is given by the product of the measurement time ($N\Delta t$) and the per-measurement power consumption ($P_\text{meter}$). In \eqref{Eabs}, $\bm{\theta}^{(n)}$ represents the phase shift configuration adopted in the $n-$th measurement interval, while $\bm{\theta}^\star$ is the phase shift configuration that is adopted after the exploration. Obviously $N\Delta t\le t_2-t_1$ is required. Moreover, observe that not only exploration consumes valuable energy resources ($N\Delta tP_\text{meter}$) but also low-performance phase shift configurations may be often tested. Therefore, it is desirable reaching to a proper solution $\bm{\theta}^\star$ with relatively few measurements $N$ to limit the performance degradation impact of the exploration phase. On the other hand, $\Delta t$ needs to be sufficiently large so the waveform-dependent random fluctuations in the measurements are averaged out. This is, $E_{t_1+(n-1)\Delta t}^{t_1+n\Delta t}(\bm{\theta}^{(n)})$ is an accurate estimate of $E_{t_1}^{t_2}(\bm{\theta}^{(n)})$, $\forall n=1,2,\cdots,N$, as considered hereinafter.
\subsection{Phase Shifters Modeling}
\begin{figure}[t!]
	\centering
	\includegraphics[width=0.49\textwidth]{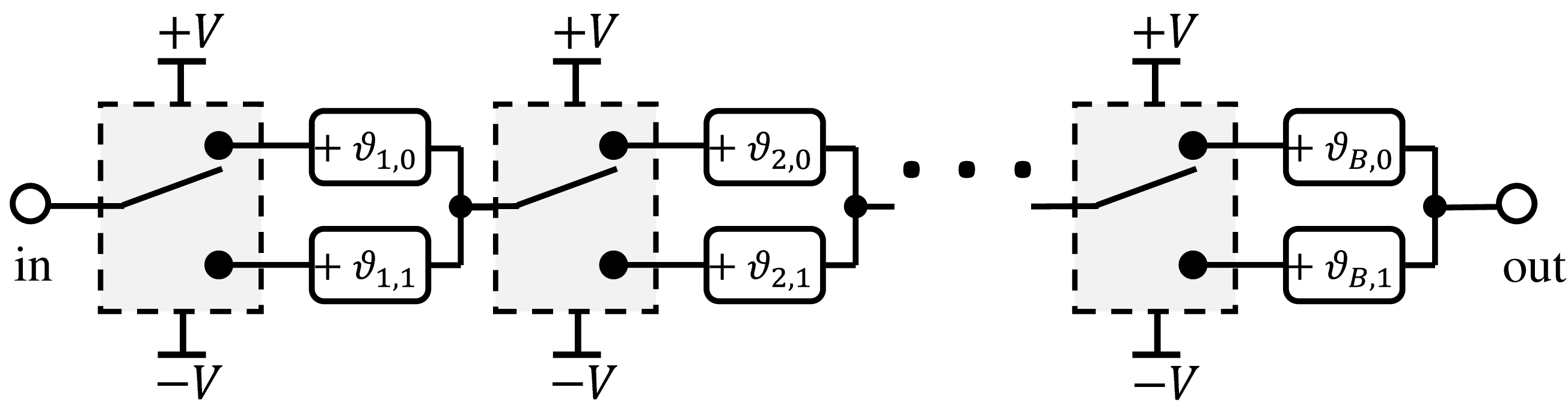}
	\vspace{-4mm}
	\caption{Digital adjustable phase shifter with $B$-bit resolution.}
	\label{Fig3}
	\vspace{-2mm}
\end{figure}

Active phase shifters implementations with continuously adjustable phase are not an option as they require power consuming and costly digital to analog converters to generate the analog control voltages \cite{Ellinger.2010}. Instead,  discretely adjustable passive phase shifters are needed. Specifically, we opt for switched-type phase shifters using two-state ($+V\mapsto$`high' and $-V\mapsto$`low') control voltages as illustrated in Fig.~\ref{Fig3}. The idea is to switch between different phase paths that can be realized, e.g., with lumped high-pass/low-pass structures \cite{Ellinger.2010}. With $B$ staged paths, i.e., a $B$-bit resolution circuit,  $|\Theta|=2^B$ different phase shifts can be realized, while $2B$ digital control voltages and maybe also contact pads are required for this task.\footnote{The number of control lines and contact pads can be reduced by applying a serializer circuit, which may become specially attractive at low-to-moderate frequencies to avoid large form-factor implementations \cite{Ellinger.2008}.}

The phase shifter architecture illustrated in Fig.~\ref{Fig3} is connected to $M-1$ antennas, i.e., all but the reference antenna. Without loss of generality, let's assume the first antenna as the reference antenna, then,
\begin{align}
\delta_j&=\left\{\begin{array}{ll}
1,& \text{if }j=1,\\
\delta_0^B, & \text{if }j\in\{2,\cdots,M\},
\end{array}\right.\\
E_{\text{phase}}&=(M-1)BP_0,\label{Ephase}
\end{align}
where $\delta_0$ and $P_0$ denote respectively the insertion loss and power consumption per bit of resolution \cite{Ellinger.2008}. Observe that the phase shift experienced by the signal impinging the $j-$th (${j\ge 2}$) antenna is given by $\theta_j=\sum_{k=1}^B\vartheta_{k,b_{j,k}}$, where $b_{j,k}\in\{0,1\}$ is the bit controlling the $k-$th phase shifting stage of the $j-$th antenna branch. 
\section{Dynamic RF Combining Schemes}\label{S5}
\subsection{Brute Force}
The brute force (BF) technique requires testing all possible permutations (with repetitions) in which $M-1$ elements can be drawn from $\Theta$. Thus,
\begin{align}
N = |\Theta|^{M-1}. \label{Nbf} 
\end{align}
Therefore $|\Theta|\le (\Delta t)^{-1/(M-1)}$ is required for feasibility. When such condition holds, the BF technique attains the optimum phase shift configuration $\bm{\theta}^\text{opt}$ at the end of the longest-possible exploration phase.
\subsection{Sequential Testing}\label{seq}
The proposed sequential testing (ST) technique requires that each (but the reference) antenna can be selectively disconnected from the rectifying process. Thus, $M-1$ circuit switching mechanisms must be implemented. ST works as follows: 
\begin{enumerate}
	\item start with only the reference antenna connected to the rectifier; 
	\item pick one antenna branch (and corresponding phase shifter) from the set of those that have not been configured yet. Declare this  antenna branch as active;
	\item  the energy meter measures the harvested power for every phase shift configuration, i.e., $\forall \theta\in\Theta$,  at the active phase shifter, and adopts the optimum phase configuration;
	\item end the procedure if  all phase shifters have been already configured, otherwise go to 2).
\end{enumerate}
Each time the optimization occurs over a single phase shifter. Therefore,
\begin{align}
N &= (M-1)|\Theta|, \label{Nst}
\end{align}
and $|\Theta|\le 1/(\Delta t (M-1))$ is required for feasibility. Moreover, the energy consumption of the phase shifting circuitry needs to account for the incorporated switching mechanism, thus, \eqref{Ephase} can be reformulated for ST as
\begin{align}
E_{\text{phase}}&=(M-1)BP_0 +  \frac{M(M-1)}{2}\Delta t|\Theta| P_\text{switch}\nonumber\\
&\simeq \big(B+\tfrac{1}{2}M\Delta t|\Theta|\big)(M-1)P_0.\label{Ephase2}
\end{align}
Here, $P_\text{switch}$ denotes the power consumed in the per-antenna switching mechanism in the measurement phase. The last line in \eqref{Ephase2} comes from using $P_\text{switch}\simeq P_0$, which holds in practice as both are of the same order.

Finally, observe that for asymptotically large phase shift sets, i.e., $|\Theta|\rightarrow\infty$, the ST technique attains the global optimum performance. This is because each phase shifter is able to remove completely the incoming signal phase offset with respect to the reference antenna, which leads to the best-possible performance. 
\subsection{Codebook Based Configuration}
A codebook based (CB) configuration lies on selecting directly the phase shifts at all branches, i.e., the codeword, from a pre-defined codebook. All the codewords are tested during the exploration phase, and the one providing the best performance is selected. A popular codebook construction is based on  the discrete Fourier matrix (DFT) since each codeword corresponds to a specific spatial direction, i.e., receive line-of-sight (LOS) direction. Therefore, DFT-based CB allows increasing the antenna array gain (by reducing the array beamwidth) in the most suitable spatial direction. 

For instance, let us assume a uniform linear array (ULA) with half-wavelength spaced antenna elements
to ease our exposition. Now, the phase shifting configurations that \textit{efficiently} sweeps the entire angular domain is determined by the collection of columns (or rows), excluding always the first row (or column), of an $M \times M$ DFT matrix as follows
\begin{align}
\begin{matrix}
\includegraphics[width=0.36\textwidth]{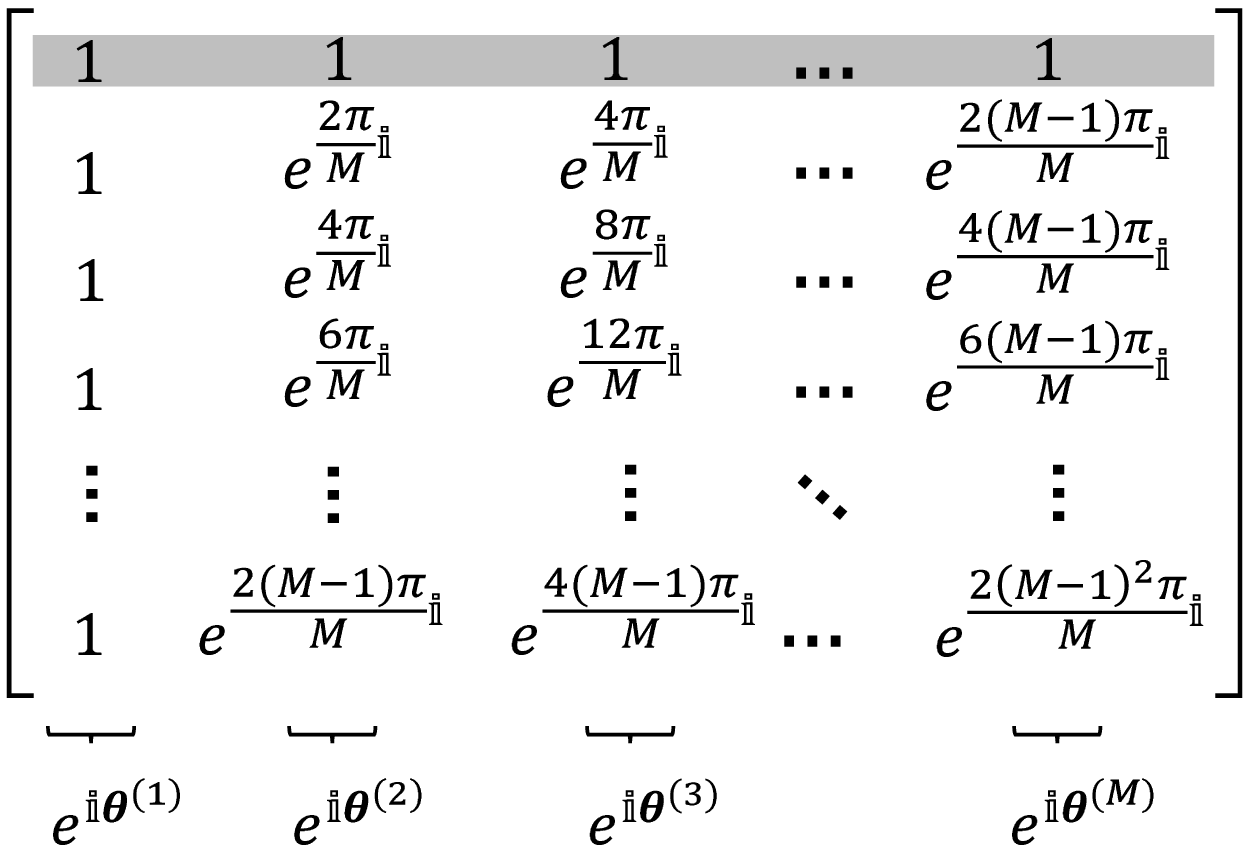}
\end{matrix}\label{matrix}
\end{align}
Observe that testing all the spatial directions requires a phase shifting set given by $\Theta=\frac{2\pi}{M}\times\big\{0,1,2,\cdots,M-1\big\}$, for which $|\Theta|=M$.\footnote{With a smaller set, the antenna array can still sweep some of the spatial directions, while more refined spatial directions can be tested with a larger set. However, a proper design must be such that $|\Theta|=M$.} This set can be realized by using the digitally adjustable phase shifter shown in Fig.~\ref{Fig3} with
\begin{align}
\vartheta_{k,l}=\left\{\begin{array}{ll}
2\pi j - \frac{\pi}{2k},& k=1,\cdots,B-1,\ l=0,1,\\
\frac{\pi}{2(B-1)}, & k=B,\ l=0,\\
0, & k=B,\ l=1,
\end{array}\right.\label{VAR}
\end{align}
and $B=\lceil\log_2 M\rceil$, where $\lceil\cdot\rceil$ is the rounding up operator. Finally, the number of measurement time intervals when using CB is given by
\begin{align}
N=|\Theta|,\label{nccnb} 
\end{align}
thus, $|\Theta|\le 1/\Delta t$ is required for feasibility.
\section{Numerical Results}\label{results}
In this section, we analyze the performance of the discussed RF combining schemes  through some numerical examples. We consider operation in the WiFi band of $2.44$ GHz, where the RF WiFi sources transmit with $p=0.1$ W and are deployed following a homogeneous Poisson point process with density $\lambda$. We set  $P_0=10$ nW, ${P_\text{meter}=80}$ nW, $\Delta t=0.5\%\times (t_2-t_1)$, and $\delta_0=-0.5$ dB \cite{Ellinger.2010}. Moreover, assume for simplicity normalized time, i.e., $t_2-t_1=1$ s, $50\%$ EH efficiency (i.e., $g(x)=0.5|x|^2$),  and quasi-static narrow-band transmissions such that $\mathbf{r}=\sum_{i=1}^{|\mathcal{S}|}\omega_i\mathbf{h}_i$, where $\omega_i$ is the power-normalized, independent of each other, signal (i.e., $\mathbb{E}[|\omega_i|^2]=1$, $\mathbb{E}[\omega_i^H\omega_j]=0\ \forall i\ne j$) transmitted by $S_i\in\mathcal{S}$, and $\mathbf{h}_i\in\mathbb{C}^{M\times 1}$ denotes the channel vector realization. Under the assumption of a ULA-equipped device, the channels are modeled using the uncorrelated Rician fading model such that $$\mathbf{h}_i\!\sim\! \sqrt{\!\frac{p\beta_i}{1\!+\!\kappa_i}}\mathcal{CN}\big(\sqrt{\kappa_i}[1,e^{-\mathbbm{i}\pi\sin\alpha_i},\cdots\!,e^{-(M\!-\!1)\mathbbm{i}\pi\sin\alpha_i}]^T,\mathbf{I}\big),$$
where  $\alpha_i$ denotes the angle with respect to the ULA's boresight direction from which the signal from $S_i$ is arriving from, $\kappa_i$ is the Rician LOS factor, and $\beta_i$ denotes the path-loss. 
We consider a log-distance path loss model with exponent $2.7$ and non-distance dependent loss of 40 dB (@ 2.44 GHz), thus $\beta_i=10^{-4}\times \max(d_i,1)^{-2.7}$, where $d_i$ is the distance between $S_i$ and the EH node, which is assumed at the origin. Since it is expected that as such distance increases, the chances of LOS decreases, we model the Rician factor as an exponentially decreasing function of $d_i$. Specifically, we set $\kappa_i=40\times e^{-\kappa_i/5}$, such that the LOS factor decreases from 
$14$ dB (@1 m) to $-4$ dB (@10 m). Finally, we generate $\Theta$ according to  \eqref{VAR}. 
\subsection{Benchmark Schemes}
The performance of the dynamic BF, ST and CB  RF combining schemes is compared with that of\!\ \footnote{Observe that a proper comparison with the performance attained with a DC or hybrid combining architecture would require modeling the rectenna's waveform-dependent non-linearities (as in \cite{Shen.2021,ShenClerckx.2021}) and heterogeneous conversion efficiencies of the rectifiers for each architecture according to the input RF range, which are out of the scope of this work.}
\subsubsection{`genie-aided'  approach (GA)}
This corresponds to the ideal RF combining, where the optimum phase shift configuration is adopted without incurring in any cost. The GA approach exploits a perfect knowledge of channels and transmit power configuration, $\{\mathbf{h}_i\}$, at each coherence interval. The performance of any practical RF combining scheme is upper-bounded by that of the GA approach. 
\subsubsection{rigid combining (RC)} This corresponds to a non-tunable RF architecture (Fig.~\ref{Fig1}b) with phase shift configuration given by  $\bm{\theta}=[0, \pi,\ 0, \pi,\cdots]^T$,
which is known to provide the 	widest main high-gain beams \cite{LopezMontejo.2020}. Moreover, we assume a fully passive implementation (i.e., no additional energy consumption sources), and no insertion losses.
\begin{figure}[t!]
	\centering
	\includegraphics[width=0.2406\textwidth]{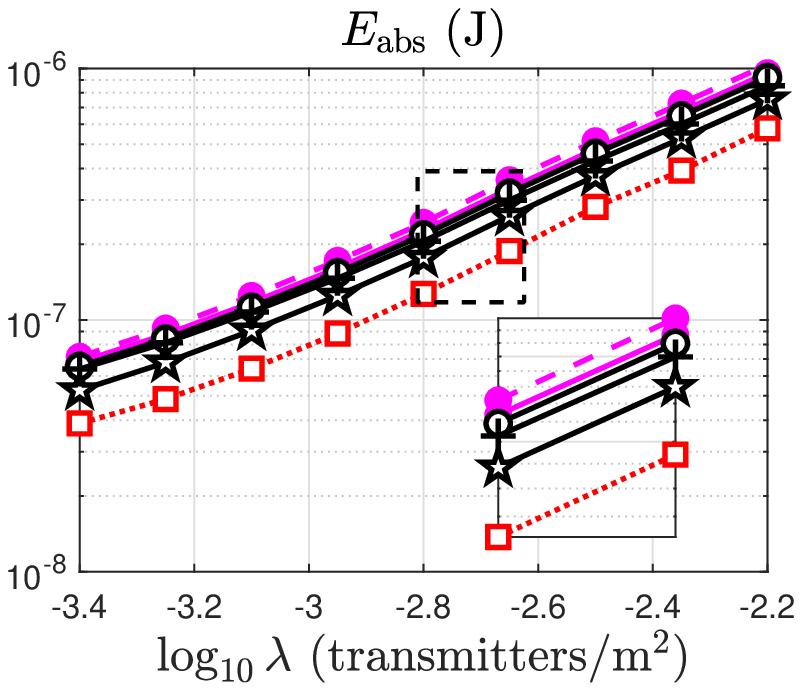}\ \ \ \ \!\!
	\includegraphics[width=0.2244\textwidth]{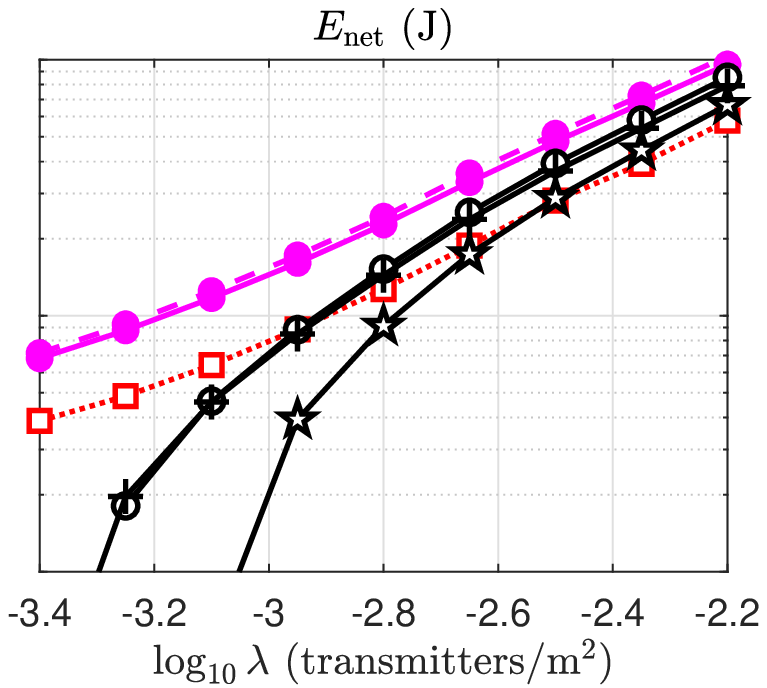}\\
	\includegraphics[width=0.48\textwidth]{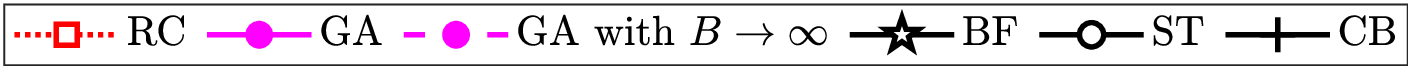}
	\caption{Absolute/net harvested energy as a function of the transmitters' deployment density. We set $M=4$ and $B=2$.}
	\label{Fig6}
	\vspace{-2mm}
\end{figure}
\subsection{On the Impact of Transmitters' Deployment Density}
Fig.~\ref{Fig6} evinces the increase of the absolute and net harvested energy with the transmitters' deployment density $\lambda$. In terms of absolute harvested energy, all the proposed dynamic mechanisms outperform RC. The situation slightly changes when accounting for the  energy consumption. Specifically, the RC approach is preferable in the region of ultra-low  EH, not so as the energy consumption of the proposed dynamic RF combining schemes becomes small with respect to the energy to be harvested.
Among the proposed mechanisms, BF performs the worst already in terms of absolute harvested energy. Observe that BF leads to the optimum phase shift configuration but it requires the longest exploration phase during which many malfunctioning configurations are tested. The latter seems to have a greater weight in the overall performance in terms of $E_\text{abs}$. The situation becomes more critical when accounting for the energy consumption since it is also the greatest among the proposed dynamic schemes, thus leading to a significant performance degradation in terms of $E_\text{net}$. Meanwhile, ST leads to a phase shift configuration that outperforms CB's in terms of absolute harvested energy despite requiring more exploration. In terms of net harvested energy, both ST and CB perform similar, being CB/ST preferable in the region of low/high EH. Finally, from the small performance gap between GA with $B=2$ and GA with $B\rightarrow\infty$, and the relative close performance of the proposed dynamic schemes, it can be deduced that a 1 or $2$-bit resolution architecture is the optimum for the adopted system configuration.
\subsection{On the Impact of the Number of Receive Antennas}
\begin{figure}[t!]
	\centering
	\includegraphics[width=0.455\textwidth]{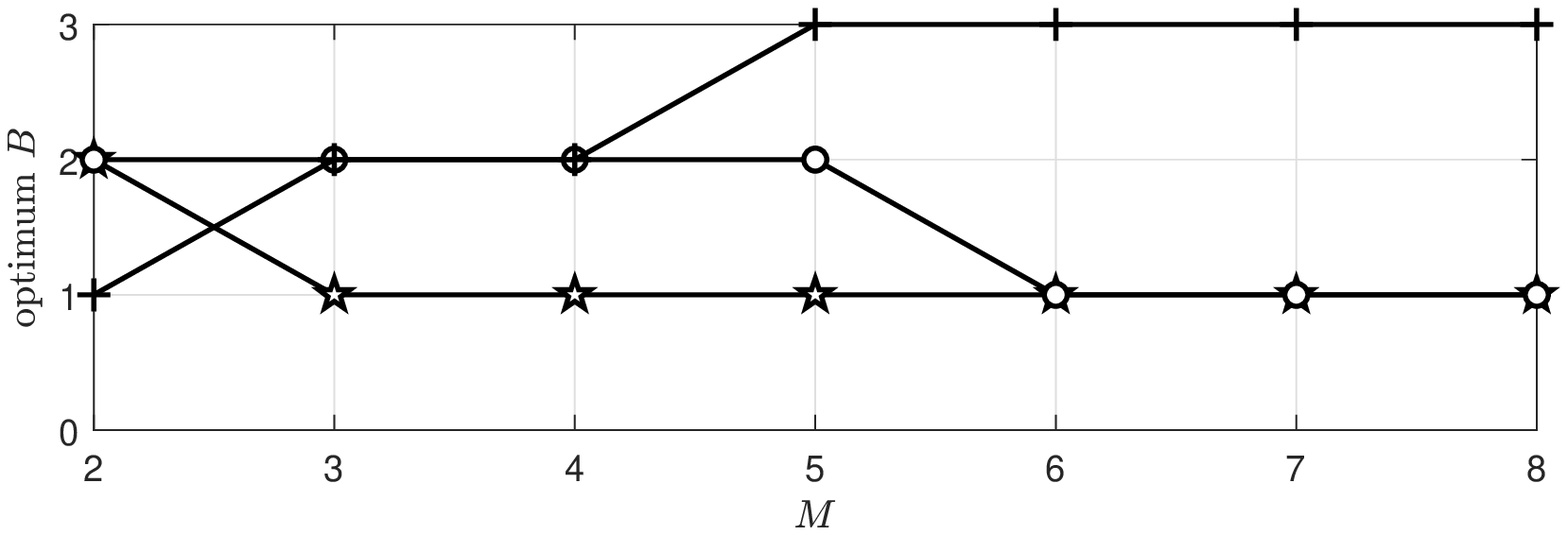}\\[0.4mm]
	\includegraphics[width=0.237\textwidth]{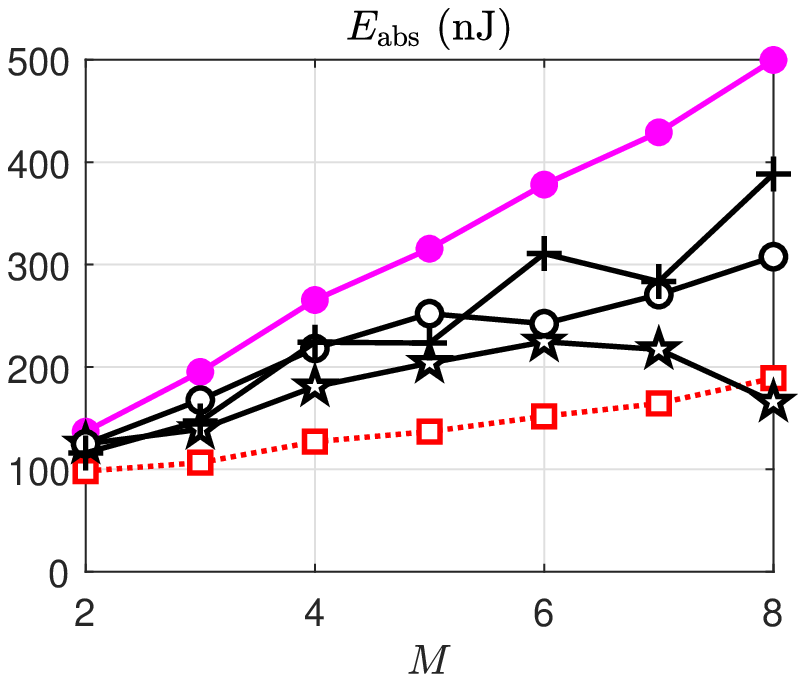}\ \ \ \ \!
	\includegraphics[width=0.222\textwidth]{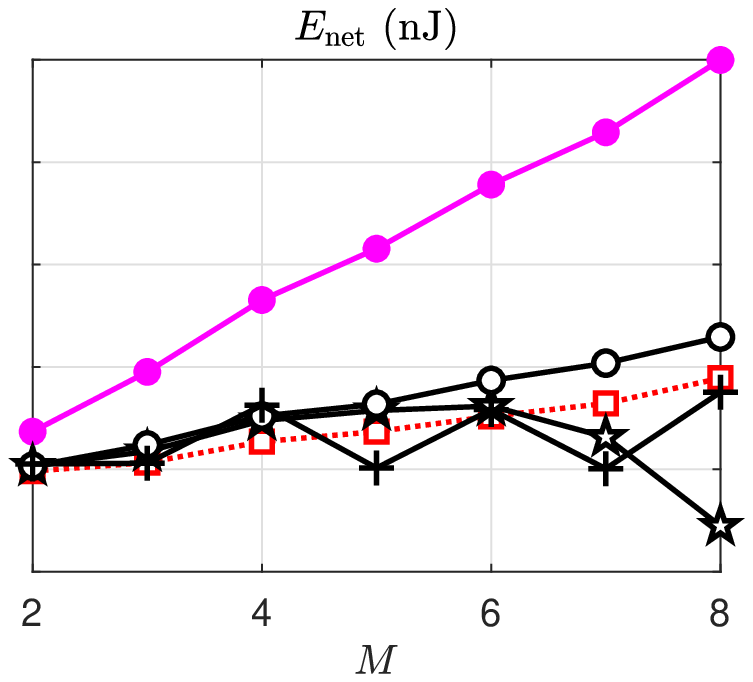}\\[0.6mm]
	\ \ \ 	\includegraphics[width=0.42\textwidth]{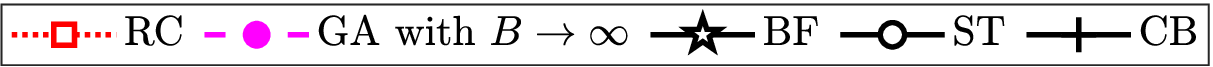}
	\caption{Optimum bit resolution (top) and corresponding absolute (bottom-left) and net (bottom-right) harvested energy  as a function of the number of receive antennas. The optimum configuration is established in terms of net harvested energy. We set $\lambda=10^{-2.8}$ transmitters/$\mathrm{m}^2$.}
	\label{Fig8}
	\vspace{-2mm}
\end{figure}

Fig.~\ref{Fig8} (top) shows the optimum, i.e., the one that provides the greatest net harvested energy, bit resolution for BF and ST as a function of the number of receive antennas $M$. Meanwhile, the corresponding absolute/net harvested energy, also for the benchmark schemes, is illustrated in Fig.~\ref{Fig8} (bottom). Observe that the DFT-based CB requires  $B=\lceil\log_2 M \rceil$, while the RC mechanism does not require any digitally tunable phase shifter. Differently from ST, for which a larger $M$ is always preferable, there is an optimum $M$ when operating with BF. This is because the EH capability after the exploration phase increases linearly with $M$ for all the schemes, while the length of the exploration phase and corresponding energy consumption of BF (ST) increases exponentially (linearly) with $M$. In general, the bit-resolution of BF is always lower than that of ST. The performance of CB with an even $M$ follows the same trend as ST's, and, in fact, the net performance gap with respect to RC slightly increases with $M$. Meanwhile, CB experiences some performance losses when $M$ is odd. The reasons are three-fold: 
i) the DFT matrix includes the codeword $[0,\pi,0,\pi,\cdots]^T$, which provides the widest receive beam (the one used by RC), only when $M$ is even. Every single codeword that can be generated when $M$ is odd is significantly less wide, thus, leading to significant performance degradation;
ii) when $M$ is not a power of two, the bit resolution cannot be fully exploited by a DFT-based CB for performance improvements; and
iii) since $B$ is a non-smooth non-decreasing function of $M$ under CB, the energy consumption  increases non-smoothly with $M$, thus heterogeneously affecting the net harvested energy.

Observe that the performance gap between the practical phase shifts' exploration mechanisms and the (ideal) genie-aided approach with infinite bit-resolution increases with $M$. This suggests that modeling the diverse impairments experienced by practical RF combining mechanisms is fundamental for appropriate performance assessing and design strategies.
\section{Conclusion}\label{conclusions}
In this letter, we introduced a dynamic RF combining architecture for ambient RF EH and three simple phase shifts' exploration mechanisms:  BF, ST and CB. We evaluated their required extra energy consumption and performance degradation due to exploration overhead, phase shifters' finite resolution and  insertion losses.  It was shown that BF demands the longest exploration and the highest power consumption, while CB requires the highest-resolution phase shifters when implemented in devices with more than two antennas, thus tipping the scales in favor of ST. Moreover, the performance gains of ST over a rigid RF combining scheme increase with the number of receive antennas and energy transmitters' deployment density. In a subsequent work, we may i) investigate suitable low-cost ML mechanisms for dynamic RF combining, and ii) perform a thoroughly comparison with DC and state-of-the-art hybrid combining schemes, for which the rectenna's EH non-linearities would be considered. 
\bibliographystyle{IEEEtran}
\bibliography{IEEEabrv,references}
\end{document}